%% file: spawc_2022_arxiv.tex
\documentclass[conference,10pt,twoside,twocolumn]{IEEEtran}
\IEEEoverridecommandlockouts

\usepackage{cite}
\usepackage[T1]{fontenc}
\usepackage{graphicx}
\usepackage{amssymb}
\usepackage{amsmath}
\usepackage{amsthm}
\usepackage{subfigure}
\usepackage{booktabs}
\usepackage{multirow}
\usepackage{microtype}
\usepackage{balance}
\usepackage{xcolor}
 
\usepackage[hidelinks]{hyperref}

\input{macros/vmr-symbols-vecbold}
\input{macros/standard-macros}

\input{macros/defs}

\newcommand{\norm}[1]{\left\lVert #1 \right\rVert}

\newcommand{\PR}[1]{\ensuremath{\!\left[#1\right]}}
\newcommand{\PC}[1]{\ensuremath{\!\left(#1\right)}}
\newcommand{\chav}[1]{\ensuremath{\!\left\{#1\right\}}}

\begin{document}

\title{An Optimization-Based User Scheduling Framework for mmWave Massive MU-MIMO Systems}

\author{\IEEEauthorblockN{Victoria Palhares and Christoph Studer}\\
\textit{Department of Information Technology and Electrical Engineering, ETH Zurich, Zurich, Switzerland} \\ \textit{e-mail: palhares@iis.ee.ethz.ch and studer@ethz.ch}\\
\thanks{The work of VP and CS was supported in part by an ETH Research Grant.  The work of CS was supported in part by ComSenTer, one of six centers in JUMP, a SRC program sponsored by DARPA. The work of CS was supported in part by the U.S.\ NSF under grants CNS-1717559 and ECCS-1824379.}
\thanks{We thank Haochuan Song and Seyed Hadi Mirfarshbafan for discussions on FBS and mmWave channels, respectively. We thank Gian Marti and Sueda Taner for comments on an early version of this paper. We also thank Remcom for providing us with a license for the Wireless InSite ray-tracing software.}}

\maketitle

\begin{abstract}
We propose a novel user equipment (UE) scheduling framework for millimeter-wave (mmWave) massive multiuser (MU) multiple-input multiple-output (MIMO) wireless systems. Our framework determines (sub)sets of UEs that should transmit simultaneously in a given time slot by approximately solving a nonconvex optimization problem using forward-backward splitting. Our UE scheduling framework is flexible in the sense that it (i) supports a variety of cost functions, including post-equalization mean square error and sum rate, and (ii) enables precise control over the minimum and maximum number of resources the UEs should occupy. We demonstrate the efficacy of our framework using realistic mmWave channel vectors generated with a commercial ray-tracer. We show that our UE scheduler outperforms a range of existing scheduling methods and closely approaches the performance of an exhaustive search.
\end{abstract}

\section{Introduction}
Millimeter-wave (mmWave) communication combined with massive multiuser (MU) multiple-input multiple-output (MIMO) is expected to be a core component in sixth generation (6G) wireless  systems \cite{Swindlehurst2014}. Combining these two technologies not only promises large beamforming gains to combat the high path loss at mmWave frequencies but also enables high-bandwidth data transmission to multiple user equipments (UEs) in the same time-frequency resource. 

In order to maximize the quality-of-service (QoS) for all UEs in the network, resource allocation strategies, such as power control and UE scheduling,  are necessary. Power control mitigates the near-far problem between the UEs transmitting to an infrastructure basestation (BS). UE scheduling distributes the UEs' requests either in time/frequency or spatially, with the goal of minimizing interference in scenarios where UEs have similar channel impulse responses---this can happen in congested scenarios with many transmitting UEs in close vicinity. The literature focuses almost exclusively on greedy UE scheduling algorithms, where one UE is selected at a time to join a set of scheduled UEs \cite{Yoo2006,Choi2019}. Such approaches have difficulties scheduling UEs over multiple time slots and fail to consider the scheduling problem globally, which results in error propagation and inevitably leads to sub-optimal QoS. 
\subsection{Contributions}
We propose a novel UE scheduling framework for mmWave MU-MIMO systems that identifies subsets of UEs that should transmit simultaneously in a given time slot. We formulate a global optimization problem, which supports a variety of cost functions, including post-linear minimum mean-square error (LMMSE) equalization mean square error (MSE) and post-LMMSE equalization sum rate. Our framework also enables precise control over the minimum and maximum number of resources each UE should occupy. In order to efficiently find approximate solutions to the nonconvex UE scheduling problem, we approximately solve a relaxed version using forward-backward splitting (FBS). We use mmWave channel vectors from a commercial ray-tracer to demonstrate that our framework outperforms a range of existing algorithms in terms of bit error rate (BER) and average per-UE rates, often closely approaching the performance of an exhaustive search (ES).
\subsection{Relevant Prior Art}
UE scheduling has been  studied extensively for mmWave communication~\cite{Choi2019, Wu2017, Zhu2021, Gao2020} and massive MU-MIMO systems \cite{Lee2018a,Farsaei2019,Jiang2018,Hu2016,Yoo2006}. Many of the proposed algorithms follow a greedy approach, such as semiorthogonal UE selection (SUS)~\cite{Yoo2006}, channel structure-based scheduling (CSS) \cite{Choi2019}, greedy max-sum rate scheduling (greedy) \cite{Choi2019}, and chordal distance-based UE scheduling (chordal) \cite{Choi2019}. All of these methods iteratively select UEs to greedily maximize a predefined cost function. In stark contrast, we address the UE scheduling problem globally, i.e., we determine the transmitting UEs for all time slots jointly by solving a single optimization problem. Our framework is flexible in (i) the choice of the cost function and (ii) the constraints on the maximum/minimum number of UEs that can transmit per time slot as well as the maximum/minimum number of time slots that a UE is allowed to transmit. Furthermore, most methods proposed in the literature aim at optimizing the signal-to-interference-plus-noise ratio (SINR), chordal distance, or channel capacity \cite{Hu2016}. Our framework also allows other, practically relevant cost functions, such as the post-equalization MSE or sum rate. Finally, we note that the majority of results for mmWave systems focus on hybrid analog-digital beamforming architectures~\cite{Choi2019, Gao2020, Wu2017, Zhu2021}. In contrast, our framework considers all-digital BS architectures, which are recently gaining popularity \cite{Mirfarshbafan2020,Dutta2020}.
\subsection{Notation}
Upper case and lower case bold symbols denote matrices and vectors, respectively. We use $A_{i,j}$ for the element on the $i$th row and $j$th column of $\bA$, $\bma_j$ as the $j$th column of~$\bA$, and~$a_i$ as the $i$th element of vector $\bma$. We define $\bI_M$, $\boldsymbol{1}_{L \times M}$, and~$\boldsymbol{0}_{L \times M}$ as the $M \times M$ identity matrix, $L \times M$ all-one matrix, and $L \times M$ all-zero matrix, respectively. The superscript $(\cdot)^H$ denotes the Hermitian transpose. A diagonal matrix with $\bma$ on the main diagonal is denoted by $\text{diag} \PC{\bma}$. The Euclidean and Frobenius norms are denoted by $\vecnorm{\cdot}_2$ and $\frobnorm{ \cdot}$, respectively. Expectation is $\Ex{}{\cdot}$ and~$\overset{e}{\leq} $ represents element-wise less-than-or-equal-to. 
\section{Prerequisites}
\subsection{System Model}
We consider the uplink of a mmWave massive  MU-MIMO system in which $U$ single-antenna UEs transmit data to an all-digital BS equipped with a $B$-antenna uniform linear array (ULA). We consider a block-fading scenario with $t = 1,\dots,T$ time slots and frequency-flat channels with input-output relation
\begin{align}
\label{eq:received_vector}
\bmy \PR{t} = \bH \bms\PR{t} + \bmn \PR{t}.
\end{align}
Here,  $\bmy \PR{t}\in\opC^{B}$ is the receive vector at time slot $t$, $\bms\PR{t}  \in \opC^{U}$ contains the transmit signals from all $U$ UEs, and $\bmn\PR{t} \in \opC^{B}$ models noise whose entries are i.i.d.\ circularly-symmetric complex Gaussian with variance $N_0$. The effective channel matrix $\bH=\tilde{\bH} \matDelta$  in \fref{eq:received_vector} combines the effect of the MIMO channel matrix $\tilde{\bH} \in \opC^{B \times U}$ and the power control matrix $\matDelta = \text{diag} \PC{\delta_1,\dots,\delta_U}$, whose entries are given by~\cite{Song2021} 
\begin{align} \label{eq:power_coefficients}
\delta_u^2 &= \text{min} \chav{\|\tilde{\bmh}_u\|_2^2,10^{\frac{\eta}{10}} \text{min}_{u'=1,\dots,U} \|\tilde{\bmh}_{u'}\|_2^2}\!/\|\tilde{\bmh}_u\|_2^2 . 
\end{align}
In \fref{eq:power_coefficients}, $\tilde{\bmh}_u$ stands for the $u$th column of $\tilde{\bH}$ and $\eta\geq0$ determines the maximum dynamic range between the weakest and strongest UE receive power in decibel (dB).
\subsection{UE Scheduling}
To formalize the UE scheduling problem, we define a binary-valued UE scheduling matrix $\bC \in \chav{0,1}^{U \times T}$, where $C_{u,t}=1$ and  $C_{u,t}=0$ indicate that UE $u$ during time slot $t$ is active and inactive, respectively. Furthermore, we define a diagonal mask matrix $\bD_{\bC}\PR{t}= \text{diag}\PC{\bmc_t} \in \chav{0,1}^{U \times U}$, where~$\bmc_t$ is the $t$th column of $\bC$. Through multiplication with $\bms\PR{t}$, this mask matrix describes which UEs transmit symbols  and which UEs are idle in time slot $t$. We absorb the effect of the mask matrix in the effective channel as $\bH\PR{t} = \bH \bD_{\bC}\PR{t}$, which now depends on $t$. Our goal is to develop a framework that determines a UE scheduling matrix $\bC$ by minimizing a given cost function~$F \PC{\bC}$ subject to constraints that specify the minimum and maximum number of resources that UEs are allowed to occupy.
\section{UE Scheduling Framework}
\subsection{Scheduling as an Optimization Problem}
The proposed UE scheduling framework consists of an application-specific cost function $F\PC{\bC}$ and a set of constraints that determine the resource utilization in time and space. We wish to solve the following optimization problem:
\begin{align}   \label{eq:original_problem}
\underset{\bC \in \chav{0,1}^{U \times T}}{\text{minimize}} \  F\PC{\bC} \  \text{subject to } \ \bC \in \setC_U  \cap \setC_T,
\end{align}
with the two constraint sets 
 \begin{align} 
 \setC_U &= \big\{U_{\text{min}} \boldsymbol{1}_{1 \times T} \overset{e}{\leq} \boldsymbol{1}_{1 \times U}\bC \overset{e}{\leq} U_{\text{max}} \boldsymbol{1}_{1 \times T}\big\} \label{eq:set_1} \\
 \setC_T  &= \big\{T_{\text{min}} \boldsymbol{1}_{U \times 1} \overset{e}{\leq} \bC\boldsymbol{1}_{T \times 1} \overset{e}{\leq} T_{\text{max}}  \boldsymbol{1}_{U \times 1}\big\}.  \label{eq:set_2}
\end{align}
The set $\setC_U$ in \fref{eq:set_1} determines the minimum $U_{\text{min}}$ and maximum~$U_{\text{max}}$ number of UEs allowed to transmit simultaneously per time slot; the set $\setC_T$ in \fref{eq:set_2} determines the minimum~$T_{\text{min}}$ and maximum~$T_{\text{max}}$ number of time slots each UE is allowed to transmit. Due to the discrete nature of the UE scheduling matrix $\bC$, the problem in \fref{eq:original_problem} is of combinatorial nature. For example, in a scenario with a total number of $U=32$ UEs with $16$ UEs transmitting in the first time slot and the remaining $16$ UEs in the second time slot, an ES would need to test over $600$ million scheduling matrices. Evidently, approximate methods to solve the scheduling problem in \fref{eq:original_problem} are necessary. 
\subsection{Problem Relaxation}
To arrive at an efficient UE scheduling algorithm, we relax the binary-valued $\{0,1\}$ entries in $\bC$ to the continuous range~$[0,1]$ as follows: $\bC \in \PR{0,1} ^{U \times T}$. While this relaxation enables the use of computationally efficient gradient-descent-based methods, it no longer enforces that the solutions are in~$\{0,1\}$. To mitigate this issue, we augment the cost function of the relaxed optimization problem with the following regularizer, which promotes binary-valued scheduling matrices $\bC$ \cite{Castaneda2018}:
\begin{align}
R\PC{\bC} = - \sum_{t=1}^T \sum_{u=1}^U \alpha \abs{C_{u,t} - 0.5}^2.
\end{align}
Here, $\alpha \geq 0$ is a regularization parameter, where larger values enforce binary-valued solutions more strictly. By utilizing  $\tilde{F}\PC{\bC} = F\PC{\bC}  +  R\PC{\bC}$ as a new cost function, we are now able to deploy numerical optimization methods to efficiently determine binary-valued solutions to the relaxed problem in~\fref{eq:original_problem}. Since the augmented cost function is nonconvex (even if the original cost function $F(\bC)$ is convex), numerical methods may only converge to a local minimum. Thus, to improve the solution quality, we perform multiple random initializations and use the solution candidate~$\bC^*$ with the lowest cost $F\PC{\bC^*}$.
\subsection{UE Scheduling via Forward-Backward Splitting (FBS)}
To solve \fref{eq:original_problem} for the set $\bC \in \PR{0,1} ^{U \times T}$ and the augmented cost function $\tilde{F}\PC{\bC} = F\PC{\bC}  +  R\PC{\bC}$, we can use FBS~\cite{Goldstein2014}. This technique performs the following step for iterations $i=1,2,\ldots$ until a stopping-condition is met:
\begin{align} \label{eq:gradient_and_proximal}
\bC^{(i+1)} = \mathrm{prox}_{\setC_U  \cap \setC_T }\PC{\bC^{(i)} - \tau^{(i)} \nabla \tilde{F}\PC{\bC^{(i)}}\!}\!.
\end{align}
Here, the proximal operator $\mathrm{prox}_{\setC_U\cap\setC_T }(\cdot)$ is the orthogonal projection onto $\setC_U \cap \setC_T $, $\tau^{(i)}$ are suitably-chosen step sizes, and~$\nabla \tilde{F}(\cdot)$ is the gradient of $\tilde{F}\PC{\cdot}$. FBS is initialized with a matrix $\bC^{(1)} \in \PR{0,1} ^{U \times T}$ drawn uniformly at random and the entries of the matrix in the last iteration $\bC^{(I_\text{max})}$ are quantized to $\{0,1\}$ to satisfy the {constraint sets} $\setC_U$ and $\setC_T$. 
\section{Proximal Operator }
\subsection{Douglas-Rachford Splitting (DRS)}
We now outline the implementation of the proximal operator $\mathrm{prox}_{\setC_U  \cap \setC_T }(\cdot)$ in \fref{eq:gradient_and_proximal}. This operator is the orthogonal projection onto an intersection of two simplexes. To determine a solution that lies within the two sets, we utilize Douglas-Rachford splitting (DRS) \cite{Douglas1956}. DRS alternatively projects onto the sets~$\setC_U$ and $\setC_T$ until a solution matrix is found. Concretely, we solve the following optimization problem
\begin{align}
\mathrm{prox}_{\setC_U  \cap \setC_T } \PC{\bZ} = \underset{\bX}{\text{arg min} } \ \psi \PC{\bX,\bZ} + \xi\PC{\bX,\bZ},
\end{align}
with the two functions 
\begin{align}
\psi \PC{\bX,\bZ} & = \frac{\beta}{2} \frobnorm{\bX-\bZ}^2 + \setX_{\setC_U} \PC{\bX}  \label{eq:proximal_a} \\
\xi \PC{\bX,\bZ}& = \frac{\beta}{2} \frobnorm{\bX-\bZ}^2 + \setX_{\setC_T}  \PC{\bX}  \label{eq:proximal_b}.
\end{align}
Here, $\beta \geq0 $ is a tuning parameter that can be used to accelerate convergence and the constraints are enforced using indicator functions $\setX_{\setC_U}$ and $\setX_{\setC_T} $, which we define as follows:
\begin{align} \label{eq:indicator_function}
\setX_{\setC} \PC{\bX} = 
\left\{\begin{array}{ll}
0 & \bX \in \setC\\
\infty & \text{otherwise.}
\end{array}\right.
\end{align}

DRS iteratively carries out the two steps \cite{Douglas1956}
\begin{align} 
&\bV^{(k+1)} =  \mathrm{prox}_{\psi}\PC{\!\bG^{(k)}\!}\! =\underset{\bX \in \setC_U  }{\argmin} \frobnorm{\bX\!-\!\frac{\beta \bZ + \bG^{(k)}}{\beta+1}}^2 \label{eq:proximal_c}\\
&\bG^{(k+1)} = \mathrm{prox}_{\xi}\PC{2\bV^{(k+1)}\!-\!\bG^{(k)}}+\bG^{(k)}-\bV^{(k+1)}, \label{eq:proximal_d} 
\end{align}
and is initialized with $\bG^{(1)} = \boldsymbol{0}_{U \times T}$. After convergence or a maximum number of iterations, DRS outputs the matrix $\bV^{(K_\text{max})}$. The details of this algorithm will be provided in~\cite{Palhares2022}.
\subsection{Projection with Inequality Constraints}
The steps in \fref{eq:proximal_c} and \fref{eq:proximal_d} project each column onto the simplexes. Since both of the sets $\setC_U$ and $\setC_T$ are described by inequalities, we need a projection onto a generic simplex with inequality constraints. Given a vector $\bmq \in \opR^{M}$, our objective is to find a solution vector  $\bmp^* \in \setC^{M}$ closest to $\bmq$ as follows:
\begin{subequations} \label{eq:proj_with_inequality}
\begin{align}
\underset{\bmp \in \opR^{M}}{\text{minimize}} \ & \frac{1}{2} \! \norm{\bmp-\bmq}_2^2 \\
\text{subject to } &l_{\text{min}} \leq \sum_{i=1}^{M} p_i \leq l_{\text{max}}, \,\, 0 \leq p_i \leq 1, \forall i.
\end{align}
\end{subequations}
To solve this subproblem efficiently, we use a line search approach that requires sorting of $\bmq$ to determine the vector $\bmp^*$ satisfying the inequality constraints. The Karush-Kuhn-Tucker (KKT) conditions and the algorithm will be detailed in \cite{Palhares2022}.
\section{Cost Functions}
We now discuss the two cost functions considered in this work. Due to space constraints, the gradients as well as other cost functions will be discussed in future work \cite{Palhares2022}.
\subsection{Post-LMMSE Equalization MSE} \label{sec:lmmse_mse}
As our first cost function, we consider the post-LMMSE MSE. We define this cost function as
\begin{align} \label{eq:mse_cost_function}
F(\bC) = \sum_{t=1}^T \Ex{}{\vecnorm{\bD_{\bC}\PR{t}\bms\PR{t}- \bD_{\bC}\PR{t}\bW\PR{t}^H \bmy\PR{t}}_2^2}\!,
\end{align}
where $\bW\PR{t} = \bH\PR{t}\PC{\bH\PR{t}^H \bH\PR{t}+ \frac{N_0}{E_s} \bI_U}^{-1}$ is the LMMSE equalization matrix in time slot $t$ and $E_s$ is the transmit signal energy. Note that this cost function only accounts for errors from the UEs that transmit in a given time slot~$t$. The gradient for this cost function will be detailed in~\cite{Palhares2022}.
\subsection{Post-LMMSE Equalization Sum Rate}  \label{sec:lmmse_sum_rate}
As our second cost function, we consider the sum of achievable rates over all UEs and time slots after an LMMSE equalizer. We define this cost function as
\begin{align} \label{eq:sum_rate_cost_function}
F(\bC) = -\sum_{t=1}^T \sum_{u=1}^U \log_2\PC{1+\textit{SINR}_u\PC{\bmc_t} }\!,
\end{align}
where the SINR of the $u$th UE in time slot $t$ is given by
\begin{align} \label{eq:sinr_u_t}
\!\!\!\!\!\!\textit{SINR}_u\PC{\bmc_t} =  \frac{  \abs{\bmw_u\PR{t}^H \!\bmh_u\PR{t}}^2}{ \sum_{u'=1,u'\neq u}^U \abs{\bmw_u\PR{t}^H \!\bmh_{u'}\PR{t}}^2 \!\!+ \frac{N_0}{E_s} \vecnorm{\bmw_u\PR{t}}_2^2}. \!\!\!\!
\end{align}
Here, $\bmw_u\PR{t}$ is the $u$th column of $\bW\PR{t}$, and $\bmh_u\PR{t}$ is the $u$th column of the effective, masked channel matrix $\bH\PR{t} = \bH \bD_{\bC}\PR{t}$. The gradient for this cost function will be detailed in~\cite{Palhares2022}.
\section{Simulation Results}
\subsection{Simulation Setup}
\begin{figure}[tp]
\centering
\includegraphics[width=0.8\columnwidth]{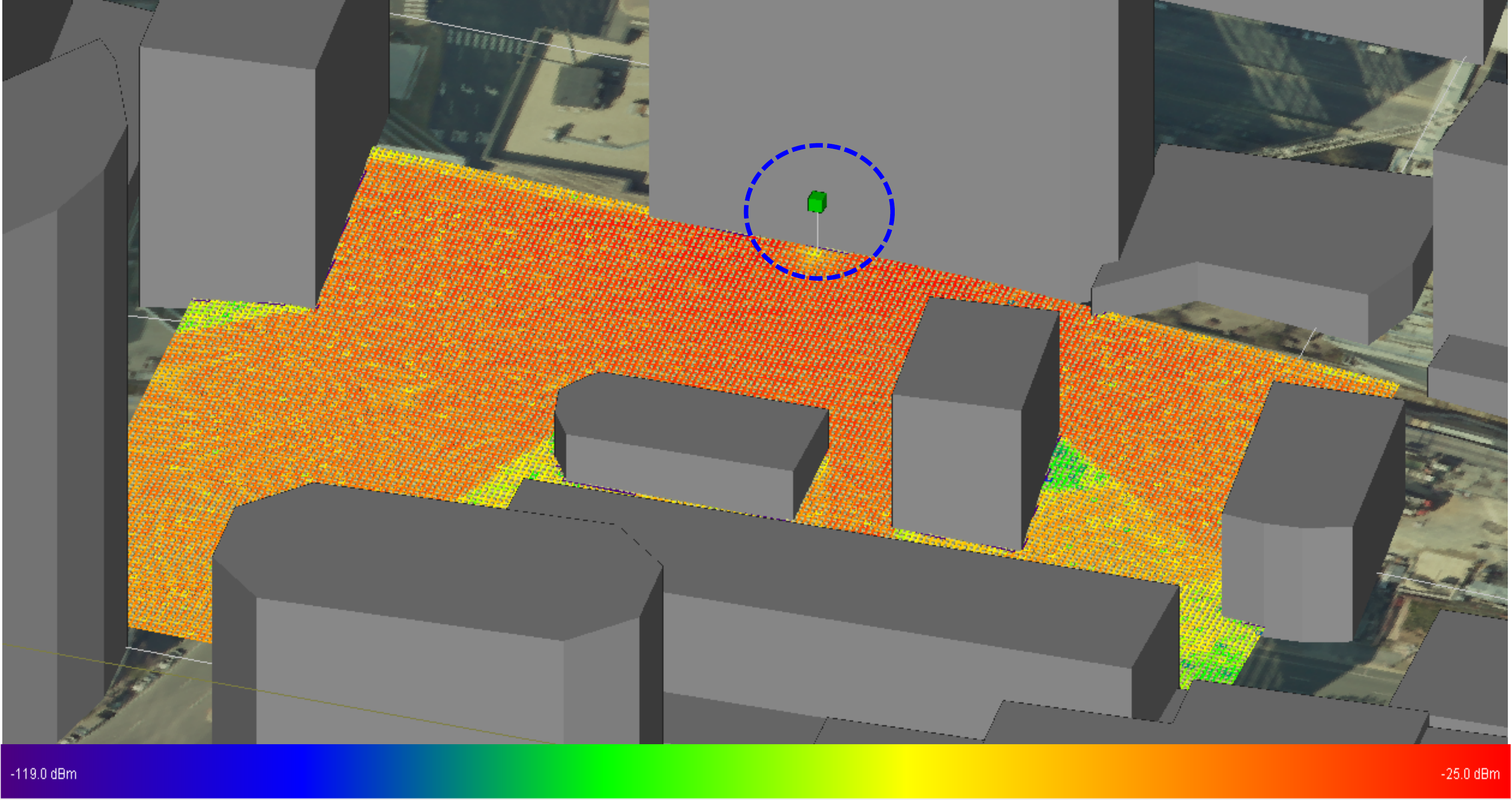}
\caption{Simulated scenario with $22,448$ UE positions and a $32$ antenna BS. Colors indicate receive power from $-119$\,dBm (blue) to $-25$\,dBm (red). }
\label{fig:remcom_scenario}
\end{figure}

We consider a mmWave massive MU-MIMO system with a carrier frequency of $60$\,GHz and a bandwidth of $100$\,MHz. At the BS, we consider an ULA with $\lambda/2$ antenna spacing. The BS and UE antennas are omnidirectional and are at a height of $10$\,m and $1.65$\,m, respectively. The UEs transmit  16-QAM symbols and the per-UE power control dynamic range is set to $\eta = 6$\,dB. We evaluate our framework using mmWave channel vectors generated with Wireless InSite~\cite{Remcom}. The channel vectors are generated for $22,448$ UE positions in an area of $109.7\,\text{m} \times 164.7\, \text{m}$. The scenario is depicted in \fref{fig:remcom_scenario}; the BS location is shown with a blue circle. To generate a channel realization, $U$ UE positions are selected uniformly and independently at random. For channel estimation, we use BEACHES~\cite{Mirfarshbafan2020}. In \fref{tbl:scenarios}, we list the four different scenarios considered in what follows and their respective number of random initializations. We set $T_{\text{min}} = T_{\text{max}} = T_S$ and $U_{\text{min}} = U_{\text{max}} = U_S$. {For the gradient step, we set $\tau^{(i)} = \tau$ to a constant stepsize.}
\begin{table}[tbp]
\centering
\caption{Evaluated Scenarios}
\label{tbl:scenarios}	
\renewcommand{\arraystretch}{1.1}
\begin{tabular}{@{}lcccccc@{}}
\toprule
Scenario& $B$ & $U$ & $T$ & $T_S$ & $U_S$ & Initializations \\ 
\midrule
S1 & $16$ & $16$ & $2$ & $1$ & $8$ & $80$ \\ 
S2 & $32$ & $32$ & $2$ & $1$ & $16$ & $10$ \\ 
S3 & $32$ & $64$ & $2$ & $1$ & $32$ & $3$ \\ 
S4 & $32$ & $64$ & $4$ & $1$ & $16 $& $3$ \\ 
\bottomrule 
\end{tabular}	
\end{table}
\subsection{Performance Metrics and Baseline Algorithms}
We consider the uncoded BER and average per-UE rate, both evaluated after utilizing an LMMSE equalizer. In our simulation results, we refer to the post-LMMSE equalization MSE as ``opt.-based MSE'' and to the post-LMMSE equalization sum rate as ``opt.-based rate.'' In order to compare our framework with baseline methods, we also simulate the ``SUS'' \cite{Yoo2006}, ``CSS'' \cite{Choi2019}, and ``greedy'' \cite{Choi2019} algorithms. For all of the mentioned baseline methods, we schedule the UEs for the first time slot and then schedule the remaining unscheduled UEs in the subsequent time slots until all of the UE requests are spread over $T$ time slots. We also consider a baseline, which picks the subsets of UEs uniformly at random (called ``random'') and a “no scheduling” baseline, in which all of the UEs are scheduled in all the time slots. For Scenario S1, we also consider an ES, which tests every possible scheduling matrix $\bC$ and selects the one that minimizes the given cost function. 
\subsection{Simulation Results}
\begin{figure}[tp]
\centering
\subfigure[BER]{\includegraphics[height=4cm]{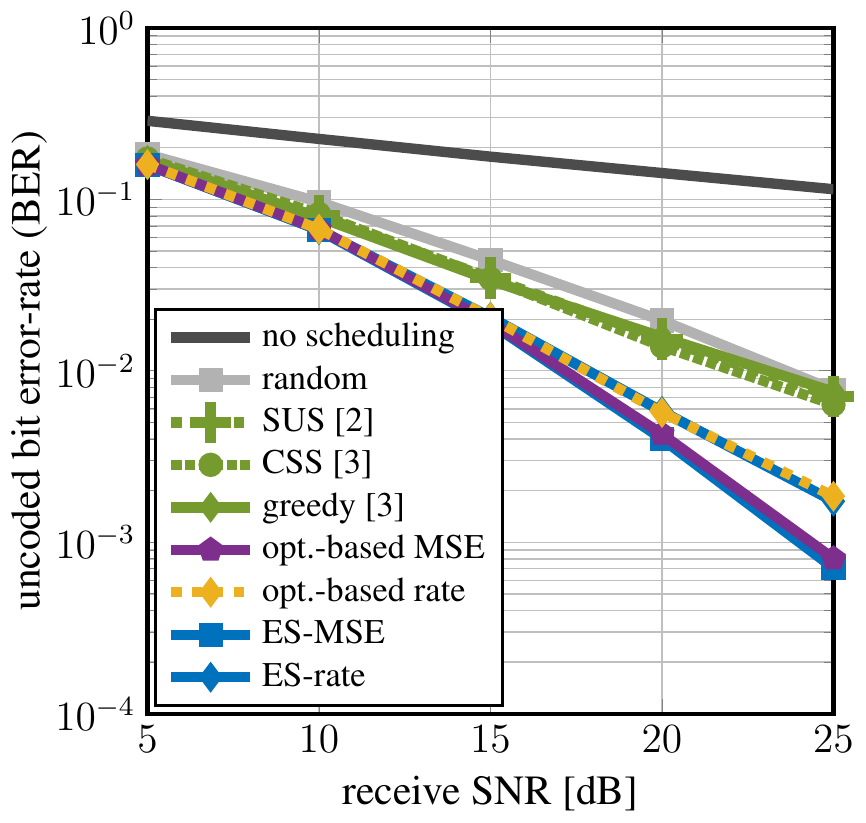}\label{fig:ber_16_16}}
\hfill
\subfigure[Average per-UE rate]{\includegraphics[height=4cm]{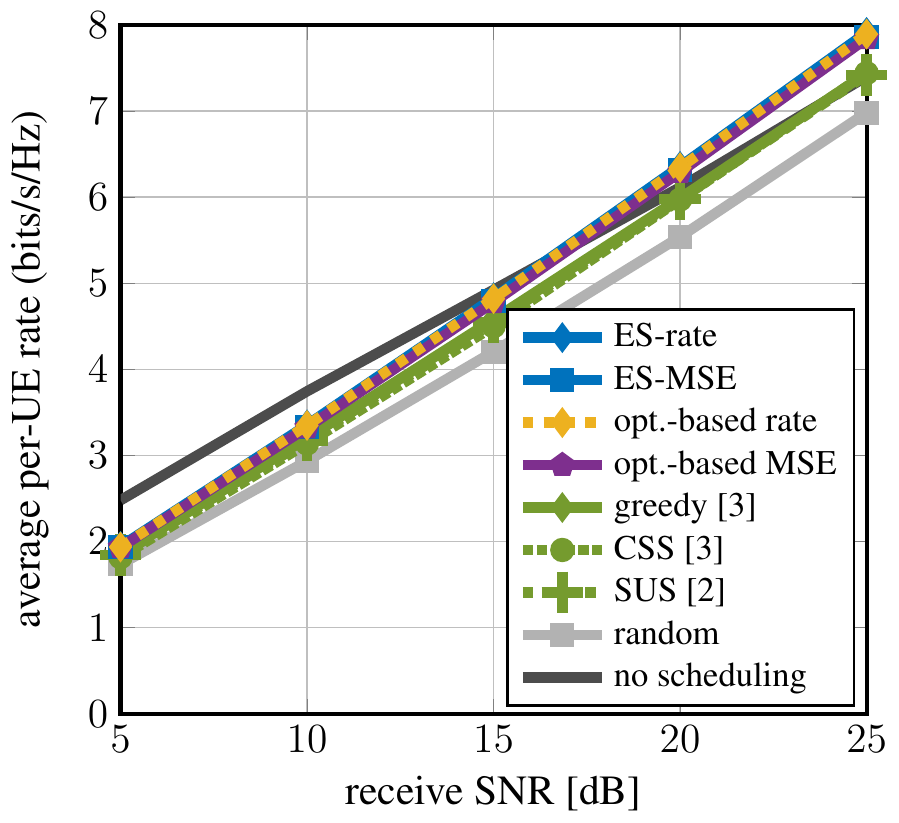}\label{fig:rate_16_16}}
\caption{(a) BER and (b) average per-UE rate for Scenario S1 with $10^2$ channel realizations and $10^5$ transmissions. The proposed UE scheduling algorithm is comparable to an ES for the MSE and sum rate criteria, in terms of BER and average per-UE rate. Our methods can reach a BER of less than $0.1$\% at a SNR of $25$\,dB. Compared to baseline methods from the literature, our framework provides a gain over $4$\,dB at 1\% BER, for example. 
} \label{fig:figures_16_16}
\end{figure}

\fref{fig:figures_16_16} shows the simulation results for Scenario S1. In \fref{fig:ber_16_16}, we see that our UE scheduling framework, both with the ``opt.-based MSE'' and ``opt.-based rate'' cost functions are comparable to the optimal baselines ``ES-MSE'' and ``ES-rate,'' reaching a BER of less than 0.1\% at a signal-to-noise ratio (SNR) of $25$\,dB. In \fref{fig:rate_16_16}, we observe the same behavior: the performance of our UE scheduling framework is nearly indistinguishable from that of an ES. Compared to existing baseline algorithms, our ``opt.-based MSE'' and ``opt.-based rate'' methods achieve superior performance. For example, compared to ``CSS,'' ``opt.-based rate'' realizes a gain of over $4$\,dB at 1\% BER. Compared with the ``no scheduling,'' ``opt.-based MSE'' and ``opt.-based rate'' are superior in terms of BER for all SNR values and in terms of average per-UE rate at high SNR.

\begin{figure*}[tp]
\centering
\subfigure[]{\includegraphics[height=3.6cm]{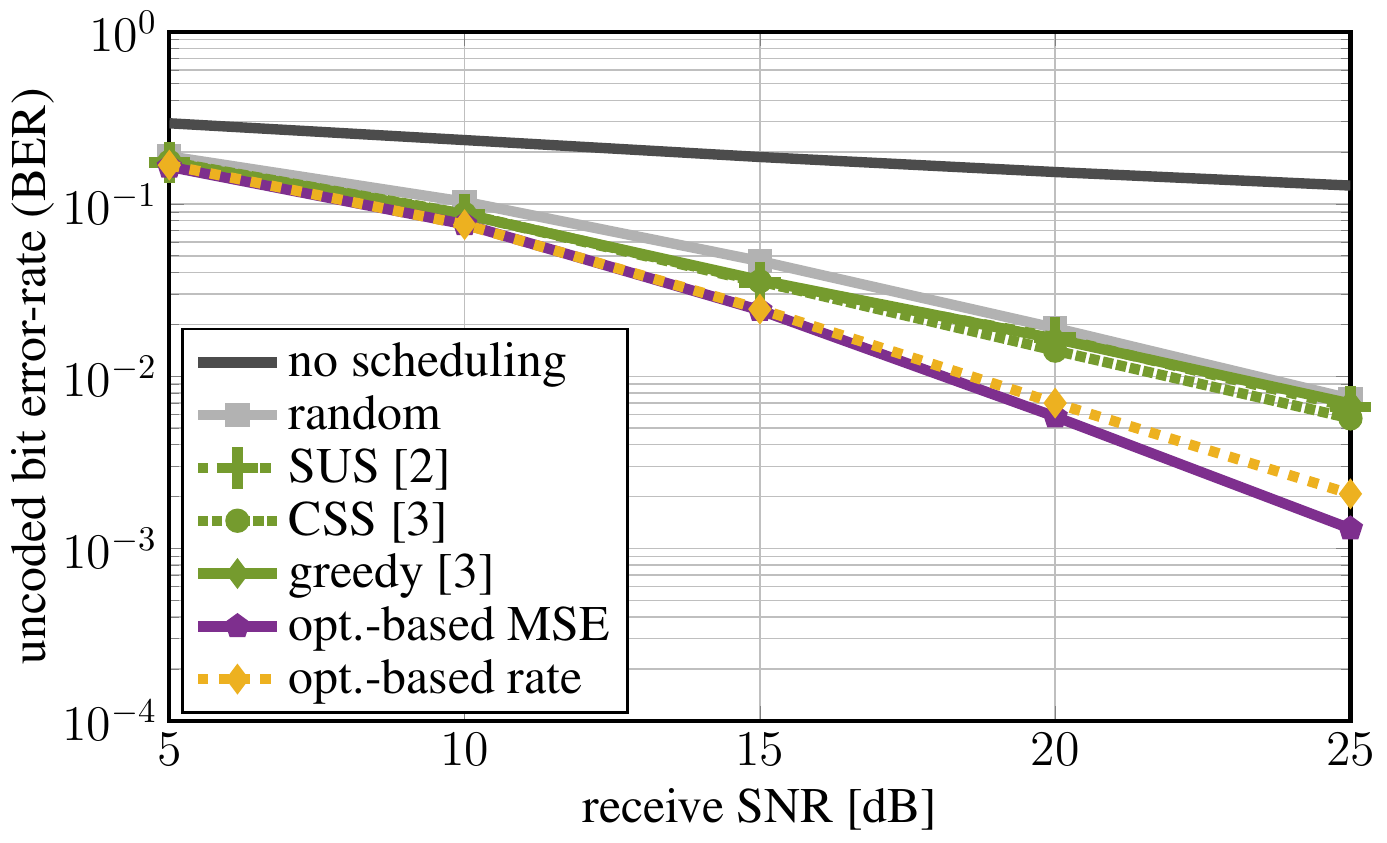}\label{fig:ber_32_32}}
\subfigure[]{\includegraphics[height=3.6cm]{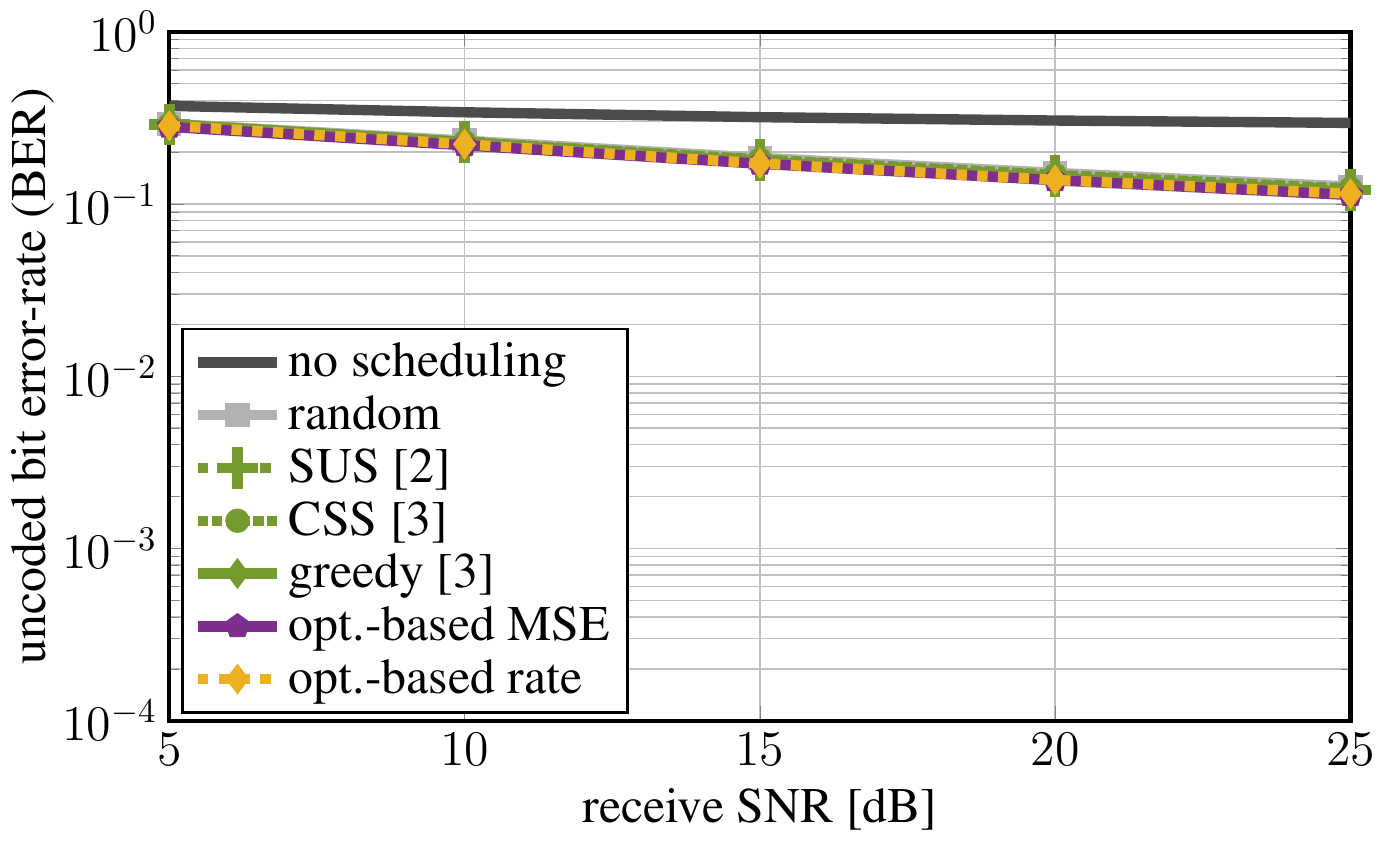}\label{fig:ber_32_64_2}}
\subfigure[]{\includegraphics[height=3.6cm]{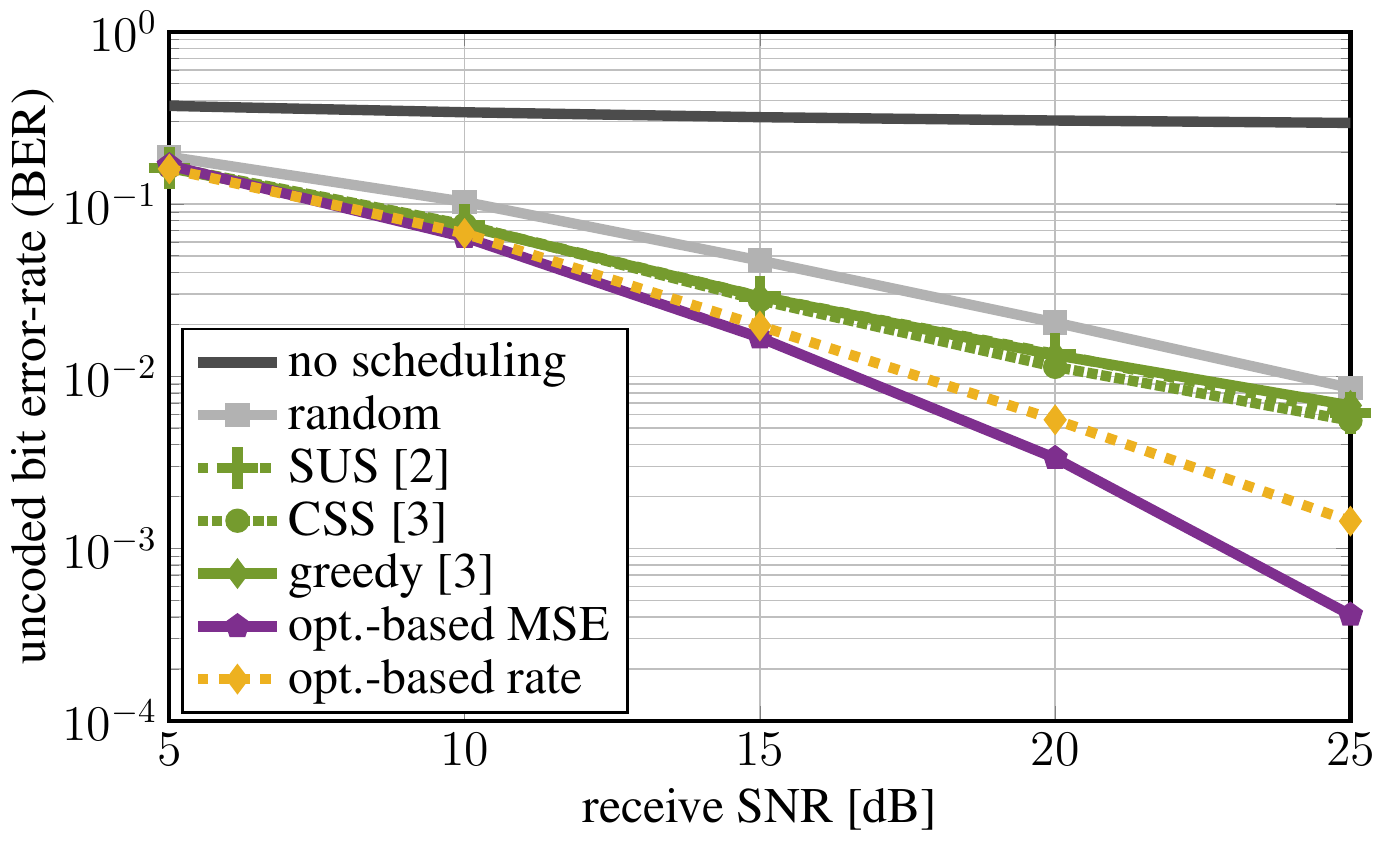}\label{fig:ber_32_64_4}}
\vspace{-0.2cm}

\subfigure[]{
\includegraphics[height=3.6cm]{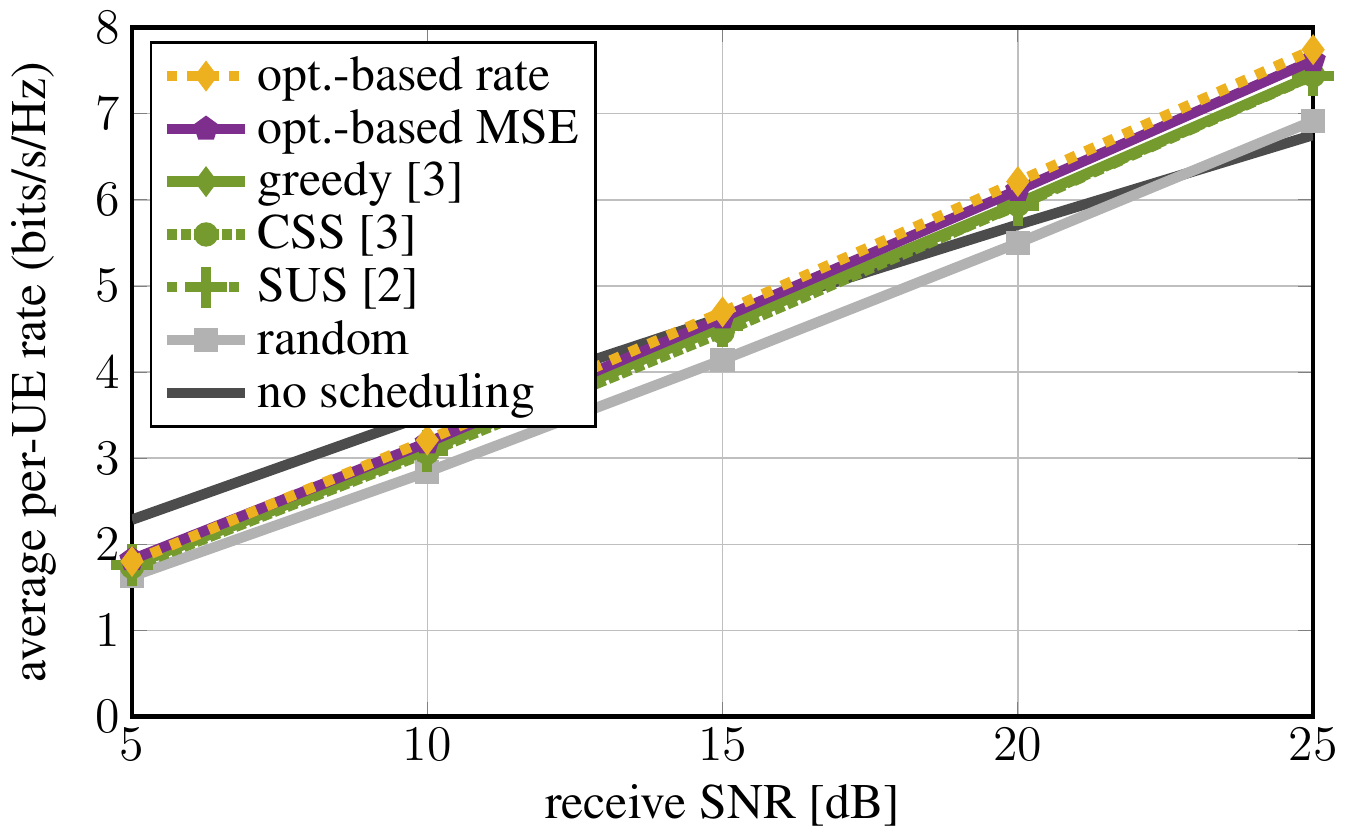}\label{fig:rate_32_32}}
\subfigure[]{
\includegraphics[height=3.6cm]{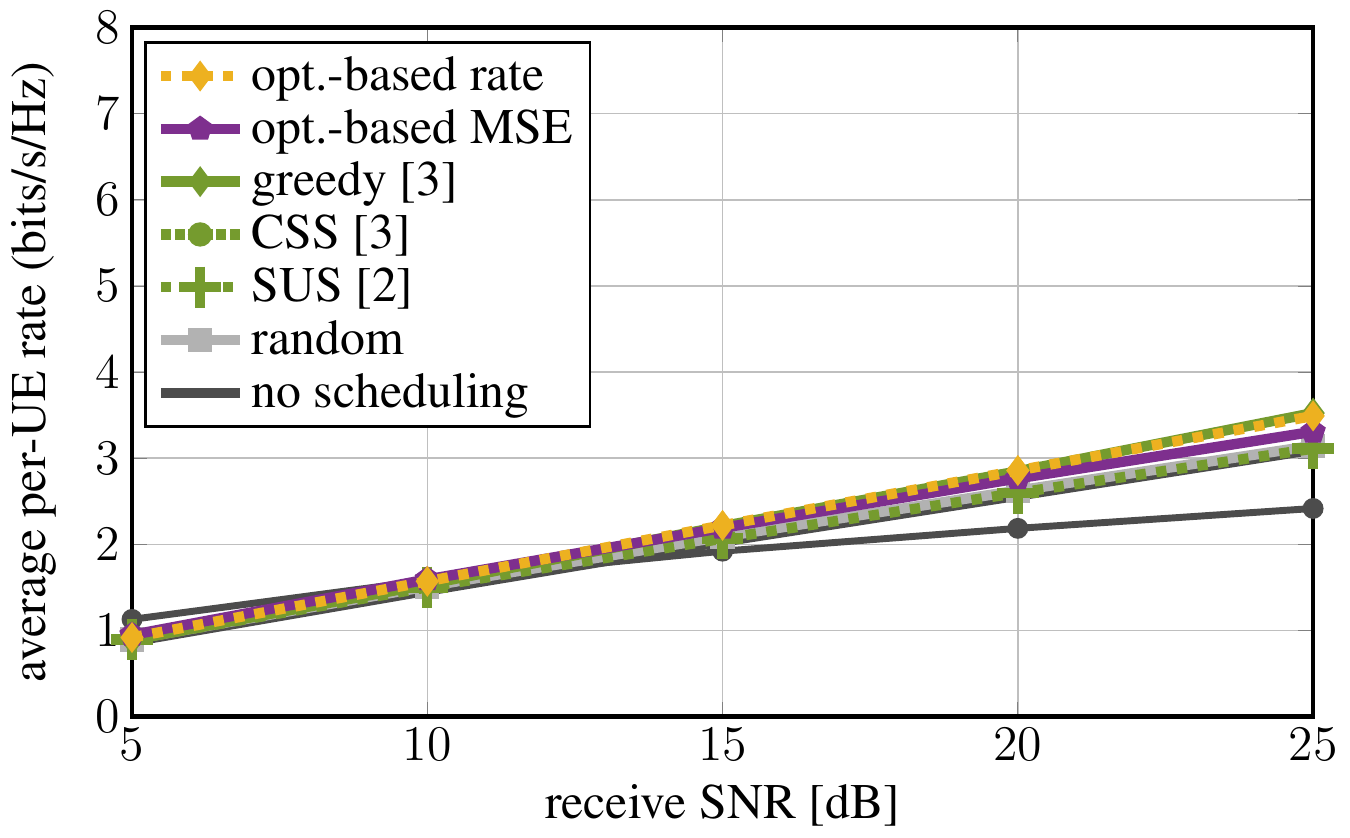}\label{fig:rate_32_64_2}}
\subfigure[]{
\includegraphics[height=3.6cm]{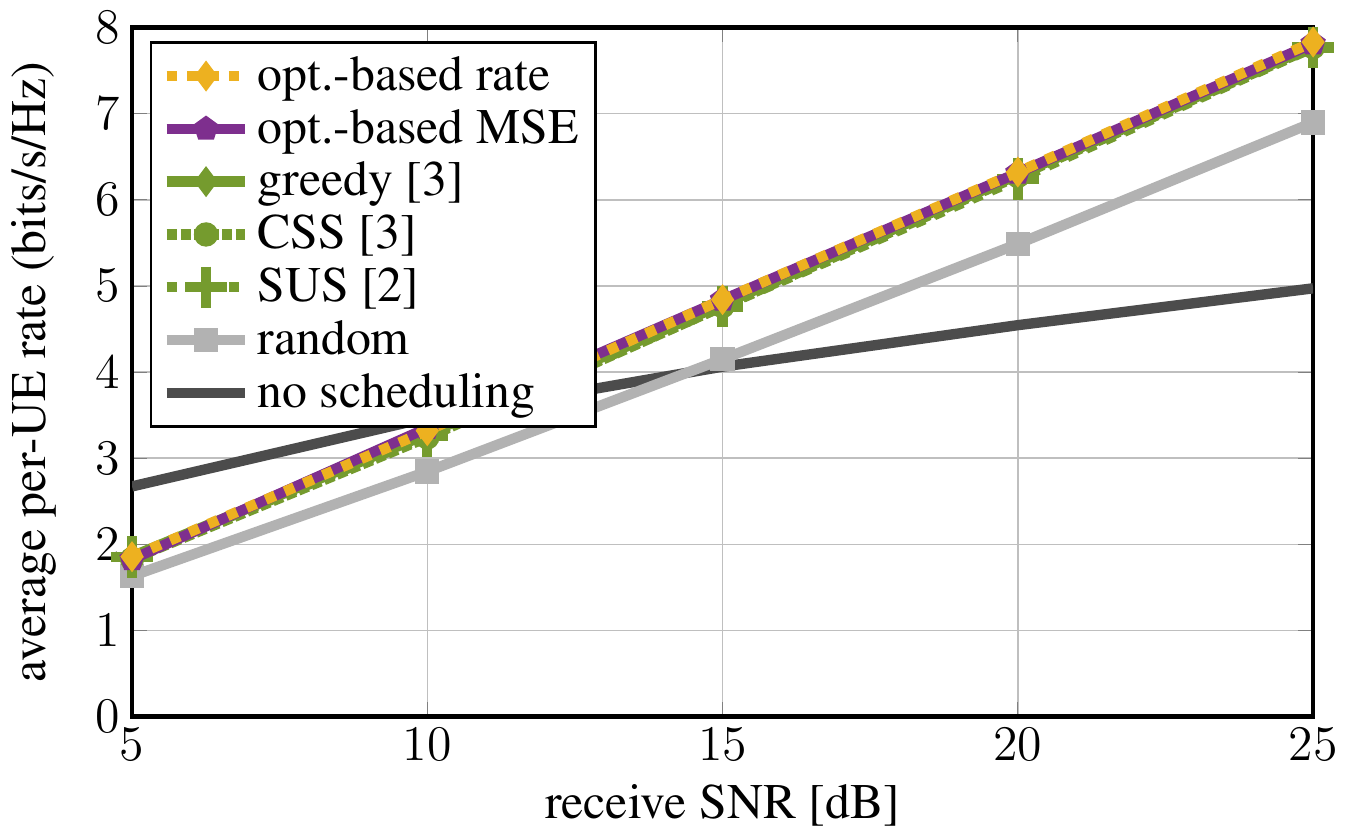}\label{fig:rate_32_64_4}}
\caption{(a), (b) and (c) are the BER performances of Scenarios S2, S3, and S4, respectively while (d), (e), and (f) are the average per-UE rates of Scenarios S2, S3, and S4, respectively. These simulations consider $10^2$ channel realizations and $10^5$ transmissions. In (a) and (c), where $U_S < B$, ``opt.-based MSE'' and ``opt.-based rate'' outperform baseline methods by $3$\,dB at 1\% BER. In (d) and (f), our proposed algorithms outperform existing  methods and the ``no scheduling'' baseline at high SNR. In (b) and (e), where $U_S = B$, even the use of scheduling does not provide acceptable QoS.}
\label{fig:figures_BER}
\end{figure*}

\fref{fig:ber_32_32}, \fref{fig:ber_32_64_2}, and \fref{fig:ber_32_64_4}  show simulation results for  Scenarios S2, S3, and S4 in terms of BER while \fref{fig:rate_32_32}, \fref{fig:rate_32_64_2}, and \fref{fig:rate_32_64_4} present the same scenarios in terms of average per-UE rate. In \fref{fig:ber_32_32}, we see that our ``opt.-based MSE'' and ``opt.-based rate'' UE scheduling algorithms have a gain of at least $3$\,dB when compared to the best baseline algorithm at 1\% BER. In terms of average per-UE rate, we see in \fref{fig:rate_32_32} that the performance of our framework outperforms existing algorithms, albeit only slightly. In \fref{fig:ber_32_64_2} and \fref{fig:rate_32_64_2}, we consider a system in which the number of scheduled UEs is equal to the number of BS antennas. Even with the use of UE scheduling, we observe that we cannot provide reasonable QoS, especially in terms of uncoded BER where it can provide a BER of approximately $10$\% at an SNR of $25$\,dB with the best scheduling method. This behavior is mainly due to the use of an LMMSE equalizer. To improve performance, one could spread the UE requests over more time slots, so that $U_S < B$, which is what we have done in Scenario~S4. We see in \fref{fig:ber_32_64_4} and \fref{fig:rate_32_64_4}, that the performance of our ``opt.-based MSE'' and ``opt.-based rate'' algorithms is superior to the baseline methods, with a gain of at least $3$\,dB at 1\% BER, for example. 

Not surprisingly, we see that utilizing the ``opt.-based MSE'' cost function provides better BER performance than ``opt.-based rate.'' Furthermore, we observe that in all scenarios, the ``no scheduling'' baseline yields extremely poor BER performance, which is due to the suboptimal LMMSE equalizer. When it comes to the average per-UE rate, as the 
UE to BS antenna ratio~$\frac{U}{B}$ or the SNR increases, the performance of ``no scheduling'' decreases. No scheduling can be beneficial at low SNR in terms of average per-UE rate, but realizing these gains would require sophisticated forward error correction strategies. 
\section{Conclusions}
We  have proposed a novel optimization-based UE scheduling framework for mmWave massive MU-MIMO systems. The framework supports a range of cost functions and constraints that specify the UE resource allocation. For a small system with $B=16$ and $U=16$, we have shown that the performance of the proposed methods is comparable to that of an exhaustive search, whereas existing baseline algorithms perform (often significantly) worse. In larger systems with $U_S < B$, we have shown that the proposed methods outperform existing methods in terms of BER and average per-UE rate. In scenarios where $U_S \geq B$, even our framework together with an LMMSE equalizer cannot achieve acceptable performance. Thus, one can schedule the UE requests over more time slots so that $U_S < B$, as it has been shown in \fref{fig:ber_32_64_4} and \fref{fig:rate_32_64_4}.

In \cite{Palhares2022}, we will introduce a range of other cost functions as well as their respective gradients. In addition, we will present the complete derivation of the projection on the simplex with inequality constraints and include a complexity analysis. 

\balance

\bibliographystyle{IEEEtran}
\bibliography{bib/VIPabbrv,bib/confs-jrnls,bib/publishers,bib/SPAWC_2022}
\balance

\end{document}

%% file: macros/vmr-symbols-vecbold.tex
\usepackage{amssymb}
\usepackage{amsfonts}
\usepackage{mathrsfs}
\usepackage{xspace}
\usepackage{bm}
\usepackage{upgreek}

\newcommand{\safemath}[2]{\newcommand{#1}{\ensuremath{#2}\xspace}}

\safemath{\bma}{\mathbf{a}}
\safemath{\bmb}{\mathbf{b}}
\safemath{\bmc}{\mathbf{c}}
\safemath{\bmd}{\mathbf{d}}
\safemath{\bme}{\mathbf{e}}
\safemath{\bmf}{\mathbf{f}}
\safemath{\bmg}{\mathbf{g}}
\safemath{\bmh}{\mathbf{h}}
\safemath{\bmi}{\mathbf{i}}
\safemath{\bmj}{\mathbf{j}}
\safemath{\bmk}{\mathbf{k}}
\safemath{\bml}{\mathbf{l}}
\safemath{\bmm}{\mathbf{m}}
\safemath{\bmn}{\mathbf{n}}
\safemath{\bmo}{\mathbf{o}}
\safemath{\bmp}{\mathbf{p}}
\safemath{\bmq}{\mathbf{q}}
\safemath{\bmr}{\mathbf{r}}
\safemath{\bms}{\mathbf{s}}
\safemath{\bmt}{\mathbf{t}}
\safemath{\bmu}{\mathbf{u}}
\safemath{\bmv}{\mathbf{v}}
\safemath{\bmw}{\mathbf{w}}
\safemath{\bmx}{\mathbf{x}}
\safemath{\bmy}{\mathbf{y}}
\safemath{\bmz}{\mathbf{z}}
\safemath{\bmzero}{\mathbf{0}}
\safemath{\bmone}{\mathbf{1}}

\bmdefine{\biad}{a}
\bmdefine{\bibd}{b}
\bmdefine{\bicd}{c}
\bmdefine{\bidd}{d}
\bmdefine{\bied}{e}
\bmdefine{\bifd}{f}
\bmdefine{\bigd}{g}
\bmdefine{\bihd}{h}
\bmdefine{\biid}{i}
\bmdefine{\bijd}{j}
\bmdefine{\bikd}{k}
\bmdefine{\bild}{l}
\bmdefine{\bimd}{m}
\bmdefine{\bind}{n}
\bmdefine{\biod}{o}
\bmdefine{\bipd}{p}
\bmdefine{\biqd}{q}
\bmdefine{\bird}{r}
\bmdefine{\bisd}{s}
\bmdefine{\bitd}{t}
\bmdefine{\biud}{u}
\bmdefine{\bivd}{v}
\bmdefine{\biwd}{w}
\bmdefine{\bixd}{x}
\bmdefine{\biyd}{y}
\bmdefine{\bizd}{z}

\bmdefine{\bixid}{\xi}
\bmdefine{\bilambdad}{\lambda}
\bmdefine{\bimud}{\mu}
\bmdefine{\bithetad}{\theta}
\bmdefine{\biphid}{\phi}
\bmdefine{\bideltad}{\delta}

\safemath{\bmia}{\biad}
\safemath{\bmib}{\bibd}
\safemath{\bmic}{\bicd}
\safemath{\bmid}{\bidd}
\safemath{\bmie}{\bied}
\safemath{\bmif}{\bifd}
\safemath{\bmig}{\bigd}
\safemath{\bmih}{\bihd}
\safemath{\bmii}{\biid}
\safemath{\bmij}{\bijd}
\safemath{\bmik}{\bikd}
\safemath{\bmil}{\bild}
\safemath{\bmim}{\bimd}
\safemath{\bmin}{\bind}
\safemath{\bmio}{\biod}
\safemath{\bmip}{\bipd}
\safemath{\bmiq}{\biqd}
\safemath{\bmir}{\bird}
\safemath{\bmis}{\bisd}
\safemath{\bmit}{\bitd}
\safemath{\bmiu}{\biud}
\safemath{\bmiv}{\bivd}
\safemath{\bmiw}{\biwd}
\safemath{\bmix}{\bixd}
\safemath{\bmiy}{\biyd}
\safemath{\bmiz}{\bizd}

\safemath{\bmxi}{\bixid}
\safemath{\bmlambda}{\bilambdad}
\safemath{\bmmu}{\bimud}
\safemath{\bmtheta}{\bithetad}
\safemath{\bmphi}{\biphid}
\safemath{\bmdelta}{\bideltad}

\safemath{\bA}{\mathbf{A}}
\safemath{\bB}{\mathbf{B}}
\safemath{\bC}{\mathbf{C}}
\safemath{\bD}{\mathbf{D}}
\safemath{\bE}{\mathbf{E}}
\safemath{\bF}{\mathbf{F}}
\safemath{\bG}{\mathbf{G}}
\safemath{\bH}{\mathbf{H}}
\safemath{\bI}{\mathbf{I}}
\safemath{\bJ}{\mathbf{J}}
\safemath{\bK}{\mathbf{K}}
\safemath{\bL}{\mathbf{L}}
\safemath{\bM}{\mathbf{M}}
\safemath{\bN}{\mathbf{N}}
\safemath{\bO}{\mathbf{O}}
\safemath{\bP}{\mathbf{P}}
\safemath{\bQ}{\mathbf{Q}}
\safemath{\bR}{\mathbf{R}}
\safemath{\bS}{\mathbf{S}}
\safemath{\bT}{\mathbf{T}}
\safemath{\bU}{\mathbf{U}}
\safemath{\bV}{\mathbf{V}}
\safemath{\bW}{\mathbf{W}}
\safemath{\bX}{\mathbf{X}}
\safemath{\bY}{\mathbf{Y}}
\safemath{\bZ}{\mathbf{Z}}

\safemath{\bZero}{\mathbf{0}}
\safemath{\bOne}{\mathbf{1}}
\safemath{\bDelta}{\mathbf{\Delta}}
\safemath{\bLambda}{\mathbf{\Lambda}}
\safemath{\bPhi}{\mathbf{\Upphi}}
\safemath{\bSigma}{\mathbf{\Upsigma}}
\safemath{\bOmega}{\mathbf{\Upomega}}
\safemath{\bTheta}{\mathbf{\Uptheta}}

\bmdefine{\biAd}{A}
\bmdefine{\biBd}{B}
\bmdefine{\biCd}{C}
\bmdefine{\biDd}{D}
\bmdefine{\biEd}{E}
\bmdefine{\biFd}{F}
\bmdefine{\biGd}{G}
\bmdefine{\biHd}{H}
\bmdefine{\biId}{I}
\bmdefine{\biJd}{J}
\bmdefine{\biKd}{K}
\bmdefine{\biLd}{L}
\bmdefine{\biMd}{M}
\bmdefine{\biOd}{N}
\bmdefine{\biPd}{O}
\bmdefine{\biQd}{P}
\bmdefine{\biRd}{R}
\bmdefine{\biSd}{S}
\bmdefine{\biTd}{T}
\bmdefine{\biUd}{U}
\bmdefine{\biVd}{V}
\bmdefine{\biWd}{W}
\bmdefine{\biXd}{X}
\bmdefine{\biYd}{Y}
\bmdefine{\biZd}{Z}

\bmdefine{\biDelta}{\Delta}
\bmdefine{\biLambda}{\Lambda}
\bmdefine{\biPhi}{\Phi}
\bmdefine{\biSigma}{\Sigma}
\bmdefine{\biOmega}{\Omega}
\bmdefine{\biTheta}{\Theta}

\safemath{\bimA}{\biAd}
\safemath{\bimB}{\biBd}
\safemath{\bimC}{\biCd}
\safemath{\bimD}{\biDd}
\safemath{\bimE}{\biEd}
\safemath{\bimF}{\biFd}
\safemath{\bimG}{\biGd}
\safemath{\bimH}{\biHd}
\safemath{\bimI}{\biId}
\safemath{\bimJ}{\biJd}
\safemath{\bimK}{\biKd}
\safemath{\bimL}{\biLd}
\safemath{\bimM}{\biMd}
\safemath{\bimN}{\biNd}
\safemath{\bimO}{\biOd}
\safemath{\bimP}{\biPd}
\safemath{\bimQ}{\biQd}
\safemath{\bimR}{\biRd}
\safemath{\bimS}{\biSd}
\safemath{\bimT}{\biTd}
\safemath{\bimU}{\biUd}
\safemath{\bimV}{\biVd}
\safemath{\bimW}{\biWd}
\safemath{\bimX}{\biXd}
\safemath{\bimY}{\biYd}
\safemath{\bimZ}{\biZd}

\safemath{\bimDelta}{\biDelta}
\safemath{\bimLambda}{\biLambda}
\safemath{\bimPhi}{\biPhi}
\safemath{\bimSigma}{\biSigma}
\safemath{\bimOmega}{\biOmega}
\safemath{\bimTheta}{\biTheta}

\safemath{\setA}{\mathcal{A}}
\safemath{\setB}{\mathcal{B}}
\safemath{\setC}{\mathcal{C}}
\safemath{\setD}{\mathcal{D}}
\safemath{\setE}{\mathcal{E}}
\safemath{\setF}{\mathcal{F}}
\safemath{\setG}{\mathcal{G}}
\safemath{\setH}{\mathcal{H}}
\safemath{\setI}{\mathcal{I}}
\safemath{\setJ}{\mathcal{J}}
\safemath{\setK}{\mathcal{K}}
\safemath{\setL}{\mathcal{L}}
\safemath{\setM}{\mathcal{M}}
\safemath{\setN}{\mathcal{N}}
\safemath{\setO}{\mathcal{O}}
\safemath{\setP}{\mathcal{P}}
\safemath{\setQ}{\mathcal{Q}}
\safemath{\setR}{\mathcal{R}}
\safemath{\setS}{\mathcal{S}}
\safemath{\setT}{\mathcal{T}}
\safemath{\setU}{\mathcal{U}}
\safemath{\setV}{\mathcal{V}}
\safemath{\setW}{\mathcal{W}}
\safemath{\setX}{\mathcal{X}}
\safemath{\setY}{\mathcal{Y}}
\safemath{\setZ}{\mathcal{Z}}
\safemath{\emptySet}{\varnothing}

\safemath{\colA}{\mathscr{A}}
\safemath{\colB}{\mathscr{B}}
\safemath{\colC}{\mathscr{C}}
\safemath{\colD}{\mathscr{D}}
\safemath{\colE}{\mathscr{E}}
\safemath{\colF}{\mathscr{F}}
\safemath{\colG}{\mathscr{G}}
\safemath{\colH}{\mathscr{H}}
\safemath{\colI}{\mathscr{I}}
\safemath{\colJ}{\mathscr{J}}
\safemath{\colK}{\mathscr{K}}
\safemath{\colL}{\mathscr{L}}
\safemath{\colM}{\mathscr{M}}
\safemath{\colN}{\mathscr{N}}
\safemath{\colO}{\mathscr{O}}
\safemath{\colP}{\mathscr{P}}
\safemath{\colQ}{\mathscr{Q}}
\safemath{\colR}{\mathscr{R}}
\safemath{\colS}{\mathscr{S}}
\safemath{\colT}{\mathscr{T}}
\safemath{\colU}{\mathscr{U}}
\safemath{\colV}{\mathscr{V}}
\safemath{\colW}{\mathscr{W}}
\safemath{\colX}{\mathscr{X}}
\safemath{\colY}{\mathscr{Y}}
\safemath{\colZ}{\mathscr{Z}}

\safemath{\opA}{\mathbb{A}}
\safemath{\opB}{\mathbb{B}}
\safemath{\opC}{\mathbb{C}}
\safemath{\opD}{\mathbb{D}}
\safemath{\opE}{\mathbb{E}}
\safemath{\opF}{\mathbb{F}}
\safemath{\opG}{\mathbb{G}}
\safemath{\opH}{\mathbb{H}}
\safemath{\opI}{\mathbb{I}}
\safemath{\opJ}{\mathbb{J}}
\safemath{\opK}{\mathbb{K}}
\safemath{\opL}{\mathbb{L}}
\safemath{\opM}{\mathbb{M}}
\safemath{\opN}{\mathbb{N}}
\safemath{\opO}{\mathbb{O}}
\safemath{\opP}{\mathbb{P}}
\safemath{\opQ}{\mathbb{Q}}
\safemath{\opR}{\mathbb{R}}
\safemath{\opS}{\mathbb{S}}
\safemath{\opT}{\mathbb{T}}
\safemath{\opU}{\mathbb{U}}
\safemath{\opV}{\mathbb{V}}
\safemath{\opW}{\mathbb{W}}
\safemath{\opX}{\mathbb{X}}
\safemath{\opY}{\mathbb{Y}}
\safemath{\opZ}{\mathbb{Z}}
\safemath{\opZero}{\mathbb{O}}
\safemath{\identityop}{\opI}

\safemath{\veca}{\bma}
\safemath{\vecb}{\bmb}
\safemath{\vecc}{\bmc}
\safemath{\vecd}{\bmd}
\safemath{\vece}{\bme}
\safemath{\vecf}{\bmf}
\safemath{\vecg}{\bmg}
\safemath{\vech}{\bmh}
\safemath{\veci}{\bmi}
\safemath{\vecj}{\bmj}
\safemath{\veck}{\bmk}
\safemath{\vecl}{\bml}
\safemath{\vecm}{\bmm}
\safemath{\vecn}{\bmn}
\safemath{\veco}{\bmo}
\safemath{\vecp}{\bmp}
\safemath{\vecq}{\bmq}
\safemath{\vecr}{\bmr}
\safemath{\vecs}{\bms}
\safemath{\vect}{\bmt}
\safemath{\vecu}{\bmu}
\safemath{\vecv}{\bmv}
\safemath{\vecw}{\bmw}
\safemath{\vecx}{\bmx}
\safemath{\vecy}{\bmy}
\safemath{\vecz}{\bmz}

\safemath{\veczero}{\bmzero}
\safemath{\vecone}{\bmone}
\safemath{\vecxi}{\bmxi}
\safemath{\veclambda}{\bmlambda}
\safemath{\vecmu}{\bmmu}
\safemath{\vectheta}{\bmtheta}
\safemath{\vecphi}{\bmphi}
\safemath{\vecdelta}{\bmdelta}

\safemath{\matA}{\bA}
\safemath{\matB}{\bB}
\safemath{\matC}{\bC}
\safemath{\matD}{\bD}
\safemath{\matE}{\bE}
\safemath{\matF}{\bF}
\safemath{\matG}{\bG}
\safemath{\matH}{\bH}
\safemath{\matI}{\bI}
\safemath{\matJ}{\bJ}
\safemath{\matK}{\bK}
\safemath{\matL}{\bL}
\safemath{\matM}{\bM}
\safemath{\matN}{\bN}
\safemath{\matO}{\bO}
\safemath{\matP}{\bP}
\safemath{\matQ}{\bQ}
\safemath{\matR}{\bR}
\safemath{\matS}{\bS}
\safemath{\matT}{\bT}
\safemath{\matU}{\bU}
\safemath{\matV}{\bV}
\safemath{\matW}{\bW}
\safemath{\matX}{\bX}
\safemath{\matY}{\bY}
\safemath{\matZ}{\bZ}
\safemath{\matzero}{\bmzero}

\safemath{\matDelta}{\bDelta}
\safemath{\matLambda}{\bLambda}
\safemath{\matPhi}{\bPhi}
\safemath{\matSigma}{\bSigma}
\safemath{\matOmega}{\bOmega}
\safemath{\matTheta}{\bTheta}

\safemath{\matidentity}{\matI}
\safemath{\matone}{\matO}

\safemath{\rnda}{A}
\safemath{\rndb}{B}
\safemath{\rndc}{C}
\safemath{\rndd}{D}
\safemath{\rnde}{E}
\safemath{\rndf}{F}
\safemath{\rndg}{G}
\safemath{\rndh}{H}
\safemath{\rndi}{I}
\safemath{\rndj}{J}
\safemath{\rndk}{K}
\safemath{\rndl}{L}
\safemath{\rndm}{M}
\safemath{\rndn}{N}
\safemath{\rndo}{O}
\safemath{\rndp}{P}
\safemath{\rndq}{Q}
\safemath{\rndr}{R}
\safemath{\rnds}{S}
\safemath{\rndt}{T}
\safemath{\rndu}{U}
\safemath{\rndv}{V}
\safemath{\rndw}{W}
\safemath{\rndx}{X}
\safemath{\rndy}{Y}
\safemath{\rndz}{Z}

\safemath{\rveca}{\bimA}
\safemath{\rvecb}{\bimB}
\safemath{\rvecc}{\bimC}
\safemath{\rvecd}{\bimD}
\safemath{\rvece}{\bimE}
\safemath{\rvecf}{\bimF}
\safemath{\rvecg}{\bimG}
\safemath{\rvech}{\bimH}
\safemath{\rveci}{\bimI}
\safemath{\rvecj}{\bimJ}
\safemath{\rveck}{\bimK}
\safemath{\rvecl}{\bimL}
\safemath{\rvecm}{\bimM}
\safemath{\rvecn}{\bimN}
\safemath{\rveco}{\bomO}
\safemath{\rvecp}{\bimP}
\safemath{\rvecq}{\bimQ}
\safemath{\rvecr}{\bimR}
\safemath{\rvecs}{\bimS}
\safemath{\rvect}{\bimT}
\safemath{\rvecu}{\bimU}
\safemath{\rvecv}{\bimV}
\safemath{\rvecw}{\bimW}
\safemath{\rvecx}{\bimX}
\safemath{\rvecy}{\bimY}
\safemath{\rvecz}{\bimZ}

\safemath{\rvecxi}{\bmxi}
\safemath{\rveclambda}{\bmlambda}
\safemath{\rvecmu}{\bmmu}
\safemath{\rvectheta}{\bmtheta}
\safemath{\rvecphi}{\bmphi}

\safemath{\rmatA}{\bimA}
\safemath{\rmatB}{\bimB}
\safemath{\rmatC}{\bimC}
\safemath{\rmatD}{\bimD}
\safemath{\rmatE}{\bimE}
\safemath{\rmatF}{\bimF}
\safemath{\rmatG}{\bimG}
\safemath{\rmatH}{\bimH}
\safemath{\rmatI}{\bimI}
\safemath{\rmatJ}{\bimJ}
\safemath{\rmatK}{\bimK}
\safemath{\rmatL}{\bimL}
\safemath{\rmatM}{\bimM}
\safemath{\rmatN}{\bimN}
\safemath{\rmatO}{\bimO}
\safemath{\rmatP}{\bimP}
\safemath{\rmatQ}{\bimQ}
\safemath{\rmatR}{\bimR}
\safemath{\rmatS}{\bimS}
\safemath{\rmatT}{\bimT}
\safemath{\rmatU}{\bimU}
\safemath{\rmatV}{\bimV}
\safemath{\rmatW}{\bimW}
\safemath{\rmatX}{\bimX}
\safemath{\rmatY}{\bimY}
\safemath{\rmatZ}{\bimZ}

\safemath{\rmatDelta}{\bimDelta}
\safemath{\rmatLambda}{\bimLambda}
\safemath{\rmatPhi}{\bimPhi}
\safemath{\rmatSigma}{\bimSigma}
\safemath{\rmatOmega}{\bimOmega}
\safemath{\rmatTheta}{\bimTheta}

%% file: macros/standard-macros.tex
\usepackage{amssymb}
\usepackage{amsfonts}
\usepackage{mathrsfs}
\usepackage{xspace}
\usepackage{bm}
\usepackage{fancyref}
\usepackage{textcomp}

\usepackage{multirow}
\usepackage{stmaryrd}

\newenvironment{textbmatrix}{	\setlength{\arraycolsep}{2.5pt}%
								\big[\begin{matrix}}{\end{matrix}\big]%
								\raisebox{0.08ex}{\vphantom{M}}}

\def\be{\begin{equation}}
\def\ee{\end{equation}}
\def\een{\nonumber \end{equation}}
\def\mat{\begin{bmatrix}}
\def\emat{\end{bmatrix}}
\def\btm{\begin{textbmatrix}}
\def\etm{\end{textbmatrix}}

\def\ba#1\ea{\begin{align}#1\end{align}}
\def\bas#1\eas{\begin{align*}#1\end{align*}}
\def\bs#1\es{\begin{split}#1\end{split}}
\def\bg#1\eg{\begin{gather}#1\end{gather}}
\def\bml#1\eml{\begin{multline}#1\end{multline}}
\def\bi#1\ei{\begin{itemize}#1\end{itemize}}

\newcommand{\lefto}{\mathopen{}\left}

\DeclareMathOperator*{\argmin}{arg\;min}		
\DeclareMathOperator{\Exop}{\opE}			

\newcommand{\Ex}[2]{\ensuremath{\Exop_{#1}\lefto[#2\right]}} 	
\newcommand{\abs}[1]{\lefto\lvert#1\right\rvert}		

\newcommand{\vecnorm}[1]{\lefto\lVert#1\right\rVert}		
\newcommand{\frobnorm}[1]{\vecnorm{#1}_{\text{F}}}	

\safemath{\dirac}{\delta}					
\safemath{\krond}{\dirac}					

\safemath{\upto}{\uparrow}
\safemath{\downto}{\downarrow}
\safemath{\iu}{j}							
\safemath{\ev}{\lambda}						
\safemath{\hilseqspace}{l^{2}}				
\newcommand{\banachfunspace}[1]{\setL^{#1}}	
\safemath{\hilfunspace}{\banachfunspace{2}}	

\safemath{\SNR}{\textit{SNR}} 				
\safemath{\PAR}{\textit{PAR}} 				
\safemath{\No}{N_0}							
\safemath{\Es}{E_s}							
\safemath{\Eb}{E_b}							
\safemath{\EbNo}{\frac{\Eb}{\No}}
\safemath{\EsNo}{\frac{\Es}{\No}}

\DeclareMathOperator{\CHop}{\ensuremath{\opH}} 
\safemath{\tvir}{\rndh_{\CHop}}				
\safemath{\tvtf}{\rndl_{\CHop}}				
\safemath{\spf}{\rnds_{\CHop}}				
\safemath{\bff}{H_{\CHop}}					

\safemath{\ircf}{r_{h}}						
\safemath{\tftvcf}{r_{s}}					
\safemath{\tfcf}{r_{l}}						
\safemath{\bfcf}{r_{H}}						

\safemath{\tcorr}{c_h}						
\safemath{\scf}{c_{s}}						
\safemath{\tfcorr}{c_{l}}					
\safemath{\fcorr}{c_{H}}						

\safemath{\mi}{I}							
\safemath{\capacity}{C}						

\safemath{\normal}{\mathcal{N}}			
\safemath{\jpg}{\mathcal{CN}}			
\safemath{\mchain}{\leftrightarrow}		

\safemath{\dB}{\,\mathrm{dB}}
\safemath{\dBm}{\,\mathrm{dBm}}
\safemath{\Hz}{\,\mathrm{Hz}}
\safemath{\kHz}{\,\mathrm{kHz}}
\safemath{\MHz}{\,\mathrm{MHz}}
\safemath{\GHz}{\,\mathrm{GHz}}
\safemath{\s}{\,\mathrm{s}}
\safemath{\ms}{\,\mathrm{ms}}
\safemath{\mus}{\,\mathrm{\text{\textmu}s}}
\safemath{\ns}{\,\mathrm{ns}}
\safemath{\ps}{\,\mathrm{ps}}
\safemath{\meter}{\,\mathrm{m}}
\safemath{\mm}{\,\mathrm{mm}}
\safemath{\cm}{\,\mathrm{cm}}
\safemath{\m}{\,\mathrm{m}}
\safemath{\W}{\,\mathrm{W}}
\safemath{\mW}{\, \mathrm{mW}}
\safemath{\J}{\,\mathrm{J}}
\safemath{\K}{\,\mathrm{K}}
\safemath{\bit}{\,\mathrm{bit}}
\safemath{\nat}{\,\mathrm{nat}}

\safemath{\define}{\triangleq}			

\safemath{\equivalent}{\sim}
\safemath{\distas}{\sim}					
\safemath{\sdiff}{\Delta}				

\safemath{\reals}{\mathbb{R}}
\safemath{\positivereals}{\reals_{+}}
\safemath{\integers}{\mathbb{Z}}
\safemath{\posint}{\integers_{+}}
\safemath{\naturals}{\mathbb{N}}
\safemath{\posnaturals}{\naturals_{+}}
\safemath{\complexset}{\mathbb{C}}
\safemath{\rationals}{\mathbb{Q}}

\newcommand*{\fancyrefapplabelprefix}{app}		
\newcommand*{\fancyrefthmlabelprefix}{thm}		
\newcommand*{\fancyreflemlabelprefix}{lem}		
\newcommand*{\fancyrefcorlabelprefix}{cor}		
\newcommand*{\fancyrefdeflabelprefix}{def}		
\newcommand*{\fancyrefproplabelprefix}{prop}		
\newcommand*{\fancyrefexmpllabelprefix}{exmpl}
\newcommand*{\fancyrefalglabelprefix}{alg}		
\newcommand*{\fancyreftbllabelprefix}{tbl}		

\frefformat{vario}{\fancyrefseclabelprefix}{Section~#1}
\frefformat{vario}{\fancyrefthmlabelprefix}{Theorem~#1}
\frefformat{vario}{\fancyreftbllabelprefix}{Table~#1}
\frefformat{vario}{\fancyreflemlabelprefix}{Lemma~#1}
\frefformat{vario}{\fancyrefcorlabelprefix}{Corollary~#1}
\frefformat{vario}{\fancyrefdeflabelprefix}{Definition~#1}
\frefformat{vario}{\fancyreffiglabelprefix}{Fig.~#1}
\frefformat{vario}{\fancyrefapplabelprefix}{Appendix~#1}
\frefformat{vario}{\fancyrefeqlabelprefix}{(#1)}
\frefformat{vario}{\fancyrefproplabelprefix}{Proposition~#1}
\frefformat{vario}{\fancyrefexmpllabelprefix}{Example~#1}
\frefformat{vario}{\fancyrefalglabelprefix}{Algorithm~#1}

%% file: macros/defs.tex
\safemath{\dictab}{[\,\dicta\,\,\dictb\,]}

\safemath{\ysig}{\bmy}
\safemath{\ysighat}{\hat{\ysig}}
\safemath{\ysigdim}{M}
\safemath{\xsig}{\bmx}
\safemath{\xsigdim}{N}
\safemath{\nx}{n_x}
\safemath{\zsig}{\bmz}
\safemath{\zsigdim}{\ysigdim}
\safemath{\rsig}{\bmr}
\safemath{\Adict}{\bA}
\safemath{\Adicttilde}{\widetilde{\Adict}}
\safemath{\Adictdim}{\outputdim\times\xsigdim}
\safemath{\avec}{\bma}
\safemath{\avectilde}{\tilde{\avec}}
\safemath{\Bdict}{\bB}
\safemath{\Bdicttilde}{\widetilde{\Bdict}}
\safemath{\Cdict}{\bC}
\safemath{\cvec}{\bmc}
\safemath{\Ddict}{\bD}
\safemath{\Ddictdim}{\ysigdim\times\xsigdim}
\safemath{\dvec}{\bmd}
\safemath{\Ddicttilde}{\widetilde{\bD}}
\safemath{\Bonb}{\bB}
\safemath{\bvec}{\bmb}
\safemath{\Bonbdim}{\ysigdim\times\ysigdim}
\safemath{\noise}{\bmn}
\safemath{\noisedim}{\ysigim}
\safemath{\err}{\bme}
\safemath{\errdim}{\ysigdim}
\safemath{\errset}{\setE}
\safemath{\nerr}{n_e}
\safemath{\delop}{\bP_\errset}
\safemath{\delopc}{\bP_{{\errset}^c}}

\safemath{\cplxi}{\imath}
\safemath{\cplxj}{\jmath}

\safemath{\dict}{\matD}
\safemath{\inputdim}{N}		
\safemath{\outputdim}{M}		
\safemath{\sparsity}{S}	
\safemath{\inputdimA}{{N_a}}	
\safemath{\inputdimB}{{N_b}}	
\safemath{\elemA}{{n_a}}	
\safemath{\elemB}{{n_b}}	
\safemath{\resA}{\matR_a}	
\safemath{\resB}{\matR_b}	
\safemath{\subD}{\matS} 
\safemath{\subA}{\matS_a} 
\safemath{\subB}{\matS_b} 
\safemath{\dicta}{\matA} 	
\safemath{\dictb}{\matB} 	
\safemath{\hollowS}{H}
\safemath{\hollowA}{H_a}
\safemath{\hollowB}{H_b}
\safemath{\cross}{Z}
\safemath{\coh}{\mu_d}			
\safemath{\coha}{\mu_a}			
\safemath{\cohb}{\mu_b}			
\safemath{\mubs}{\nu}	
\safemath{\cohm}{\mu_m} 
\safemath{\dictset}{\setD}	
\safemath{\dictsetp}{\dictset(\coh,\coha,\cohb)}	
\safemath{\dictsetgen}{\dictset_\text{gen}}
\safemath{\dictsetgenp}{\dictsetgen(\coh)}
\safemath{\dictsetonb}{\dictset_\text{onb}}
\safemath{\dictsetonbp}{\dictsetonb(\coh)}

\safemath{\leftside}{U}
\safemath{\rightsideA}{R_a}
\safemath{\rightsideB}{R_b}

\safemath{\indexS}{\setI_S} 

\safemath{\na}{n_a}			
\safemath{\nb}{n_b}			
\safemath{\coeffa}{p_i}	
\safemath{\coeffb}{q_j}	
\safemath{\seta}{\setP}		
\safemath{\setb}{\setQ}     
\safemath{\setw}{\setW}	
\safemath{\setz}{\setZ}	
\safemath{\cola}{\veca}		
\safemath{\colb}{\vecb}		
\safemath{\cold}{\vecd}		
\safemath{\inputvec}{\vecx} 	
\safemath{\error}{\vece}	
\safemath{\noiseout}{\vecz} 	
\safemath{\inputvecel}{x}
\safemath{\inputveca}{\vecx_a}
\safemath{\inputvecb}{\vecx_b}
\safemath{\outputvec}{\vecy}	
\safemath{\lambdamin}{\lambda_{\mathrm{min}}}

\safemath{\elltwo}{\ell_2}
\safemath{\ellone}{\ell_1}
\safemath{\ellzero}{\ell_0}
\safemath{\ellinf}{\ell_\infty}
\safemath{\ellinftilde}{\ell_{\widetilde\infty}}
\safemath{\licard}{Z(\coh,\coha,\cohb)}
\safemath{\xsol}{\hat{x}}
\safemath{\xbord}{x_b}		
\safemath{\xstat}{x_s}		
\safemath{\xstatLone}{\tilde{x}_s}
\safemath{\order}{\mathcal{O}} 
\safemath{\scales}{\Theta} 
\safemath{\ones}{\mathbf{1}} 
\safemath{\zeroes}{\mathbf{0}} 
\safemath{\thlone}{\kappa(\coh,\cohb)} 
\safemath{\constoneA}{\delta} 
\safemath{\constoneB}{\epsilon} 
\safemath{\nlarge}{L}				   
\safemath{\sumlarge}{S_\nlarge}
\safemath{\maxlarger}{P_\nlarge}	   
\safemath{\Pzero}{\textrm{P0}}	
\safemath{\Pone}{\textrm{P1}}
\safemath{\vecfir}{\vecw}			 
\safemath{\vecsec}{\vecz}
\safemath{\elvecfir}{w}              
\safemath{\elvecsec}{z}				 
\safemath{\nlargefir}{n}
\safemath{\normout}{\gamma}
\safemath{\auxfun}{h}
\safemath{\supp}{\textrm{supp}}

\safemath{\indexa}{\ell}
\safemath{\indexb}{r}
\safemath{\indexc}{i}
\safemath{\indexd}{j}

\safemath{\project}{P}